\documentclass[12pt]{article}
\usepackage{amsfonts} 
\newcommand{\be}{\begin{equation}}
\newcommand{\ba}{\begin{eqnarray}}
\newcommand{\ee}{\end{equation}}
\newcommand{\ea}{\end{eqnarray}}
\newcommand{\cosech} { {\rm cosech}}

\begin{document}

\title{The Scattering amplitude for Rationally extended shape invariant Eckart potentials}

\author{ Rajesh Kumar Yadav $^{a}$\footnote{e-mail address: rajeshastrophysics@gmail.com  }, 
Avinash Khare$^{b}$\footnote {e-mail address: khare@iiserpune.ac.in}
 and  {Bhabani Prasad Mandal$^{a}$}\footnote {e-mail address: bhabani.mandal@gmail.com}}
 \maketitle
{$~^a$ Department of Physics,Banaras Hindu University,Varanasi-221005, INDIA.\\ 
$~^b$Raja Ramanna Fellow, Indian Institute of Science Education and Research (IISER),Pune-411021, INDIA.}

\begin{abstract}
We consider the rationally extended exactly solvable Eckart potentials which 
exhibit extended shape invariance property. These potentials are isospectral to the conventional 
Eckart potential. The scattering amplitude for these rationally extended 
potentials is calculated analytically for the generalized 
$m$th ($m= 1,2,3,...$) case  by considering 
the asymptotic behavior of the scattering state wave functions which are 
written in terms 
of some new polynomials related to the Jacobi polynomials. As expected, 
in the $m=0$ limit, this scattering amplitude goes over to the scattering 
amplitude for the conventional Eckart potential.  
\end{abstract}

\section{Introduction}
 
The ideas of supersymmetric quantum mechanics (SQM) and
shape invariant potentials (SIP) have not only enriched our understanding
of the exactly solvable systems but have helped in substantially
increasing the list of exactly solvable potentials \cite{cks}. With the 
recent discovery \cite{eop2,eop3} of the exceptional Laguerre and Jacobi 
polynomials, new shape invariant
potentials with translation related to the radial oscillator, Scarf I 
and generalized 
P\"oschl-Teller potentials were discovered \cite{que,bqr} whose solution
is in terms of $X_{1}$ Laguerre or $X_{1}$ Jacobi polynomials. Subsequently,
Odake and Sasaki generalized this construction and for all three cases
obtained one parameter families of these  SIPs (with translation) whose 
bound state eigenfunctions
are in terms of $X_m$ EOPs \cite{hos,os}. In addition to this SQM approach,
there are several equivalent approaches such as Darboux-Crum 
transformation, Darboux-B\"acklund transformation, prepotential methods 
etc, which can also be used to get the extension of the conventional SIPs. 
However, till today, these three remain the only real potentials for which
rationally extended SIPs with unbroken SUSY
solutions are written in terms of EOPs. It is worth mentioning that all
three new SIPs  (with translation) are isospectral to the well 
known conventional SIPs. 
 
Using the concept of first order SQM but broken SUSY, recently Quesne
has been able to construct rational extensions of the  Eckart potential 
\cite{nesi}  and has shown that
in this case the SI property is no more valid, rather they  exhibit an 
unfamiliar
extended SI property in which the partner potential is obtained by 
translating both the potential parameter $A$ (as in the conventional case)
and $m$, the degree  of the polynomial arising in the denominator. 

Out of these newly extended SI potentials, the potential which is 
isospectral to the generalized 
P\"oschl-Teller (GPT) potential and rationally extended SI Eckart 
potentials have both the bound as well as the continuum spectrum.
In recent publication, we have obtained the scattering amplitude for the 
potential isospectral to the GPT and whose solutions 
are in terms of $X_m$ Jacobi polynomials \cite{rab1,rab2}. To the best 
of our knowledge the scattering amplitude for the rationally extended 
SI Eckart potentials have not been obtained so far. The purpose of this 
note is to fill this gap by calculating the s-wave scattering 
amplitude $(S_{l=0}^{m}(k))$ for this extended class of SI Eckart potentials whose 
solutions are in terms of some new type of polynomial 
$y_n(x)$ (which can be expressed in terms of combination of Jacobi 
polynomials) for general values of $m (=1,2,...,)$. As a check on our 
calculation, we also show that in the limit $m=0$, we recover the
scattering amplitude for the conventional Eckart potential.

The plan of the paper is as follows: In Sec. $2$, we 
briefly review the work of Quesne regarding the bound states 
for the rationally-extended SI Eckart Potentials which are 
isospectral to the conventional Eckart potential. The calculation of 
the scattering amplitude for these  potentials is discussed in Sec. $3$ 
while Sec. $4$ is reserved for summary and discussion.

\section{Rationally extended Eckart potentials: bound states}
In this section we briefly review the work of
Quesne \cite{nesi} regarding rationally-extended SI Eckart potentials
and its bound states.
The conventional Eckart potential is given by \cite{ks,lev} 
\be\label{ekrt}
V_{A,B}(r) = A(A-1)\cosech^2 r-2B\coth r, \qquad 0 < r < \infty\,, 
\ee 
where  $A>1$ and $B>A^2$, has a finite number of bound states.
The energy eigenvalues and the eigenfunctions are given by
\be\label{een}
E_\nu^{(A,B)} = -(A+\nu )^2 - \frac{B^2}{(A+\nu )^2}\,, 
\quad \nu = 0,1,2,....\nu_{max}\,, \sqrt{B}-A-1\le\nu_{max}<\sqrt{B}-A\,,
\ee    
and 
\be\label{ewf}
\psi_\nu^{(A,B)}(r) = (z - 1)^{-\frac{1}{2}(A+\nu
-\frac{B}{A+\nu})}(z + 1)^{-\frac{1}{2}(A+\nu+\frac{B}{A+\nu})}
P_\nu^{(-A-\nu +\frac{B}{A+\nu},-A-\nu -\frac{B}{A+\nu})}(z)\,,
\ee
where $z=\coth r$ and 
$ P_\nu^{(-A-\nu +\frac{B}{A+\nu},-A-\nu -\frac{B}{A+\nu})}(z)$ is 
the Jacobi polynomial.
The rational extension of this potentials has been done by Quesne by 
determining all possible polynomials-type, nodeless 
solutions $\phi(r)$ (see Ref.\cite{nesi})  of the Schr\"odinger equation 
\be\label{se}
-\frac{d^2\phi(r)}{dr^2}+V_{A,B}(r)\phi(r) = E\phi(r)\,,
\ee  
with the factorization energy $E<E_0^{(A,B)}= -A^2-\frac{B^2}{A^2}$.

Out of all the possible solutions of $\phi(r)$, two independent polynomial type solutions
$\phi_1(r)$ and $\phi_2(r)$ with the energy $E_1$ and $E_2$ respectively 
have been constructed. On putting some restrictions on 
the parameters $A$ and $B$, four polynomial type solutions (one 
corresponding to $\phi_1(r)$ and three corresponding to $\phi_2(r)$) 
have been obtained. Out of these four possibilities, there exist three 
acceptable polynomial-type nodeless solutions (one corresponding to 
$\phi_1(r)$ and two corresponding to $\phi_2(r)$) of the Eckart potentials.

Each of the above factorization function $(\phi(r))$ gives rise to a pair 
of partner potentials through the superpotential  $W(r)=-(\ln\phi(r))'$,
i.e.  
\ba
V^{(\pm)}(r)=W^2(r)\mp W'(r)+E\,.
\ea
One can now define the raising and lowering operators
\be\label{op}
\hat{A}^{\dagger}=-\frac{d}{dr}+W(r),\qquad  \hat{A}=\frac{d}{dr}+W(r)\,.
\ee 
The factorized Hamiltonians $H^{(+)}= \hat{A}^{\dagger}\hat{A}$ and $H^{(-)}= \hat{A}\hat{A}^{\dagger}$, can then be expressed as 
\ba
H^{(\pm)}= -\frac{d^2}{dr^2}+V^{(\pm)}(r)-E\,,
\ea
and satisfy the intertwining relations $\hat{A}H^{(+)}=H^{(-)}\hat{A}$ 
and $\hat{A}^{\dagger}H^{(-)}=H^{(+)}\hat{A}^{\dagger}$. 
As shown by Quesne, in this way the factorization functions $\phi(r)$ 
yield three partners potentials $V^{(-)}(r)$, out of which two are 
isospectral since their inverses are not normalizable, while the  
third partner has an additional bound state below the spectrum of $V^{(+)}(r)$, corresponding to its normalizable inverse.

The rationally-extended Eckart potential $V^{(-)}(r)$ with given $A$ 
and $B$ is obtained from
a conventional Eckart potential $V_{A,B}(r)$ by shifting the parameters 
$A$ as
\be\label{extd}
V^{(+)}(r)=V_{A',B}(r),\quad V^{(-)}(r) \equiv  V_{A,B,ext}(r)
 = V_{A,B}(r)+V_{A,B,rat}(r)\,,
\ee
where 
\be\label{rat}
V_{A,B,rat}(r)=2(1-z^2)\bigg[2z\frac{\dot{g}_m^{(A,B)}(z)}{g_m^{(A,B)}(z)}
-(1-z^2)\bigg({\frac{\ddot{g}_m^{(A,B)}(z)}{g_m^{(A,B)}(z)}
-\bigg (\frac{\dot{g}_m^{(A,B)}(z)}{g_m^{(A,B)}(z)}\bigg)^2}\bigg) \bigg]\,,
\ee
here dot denotes a derivative with respect to $z$.\\
By choosing $A'=A-1$, and the other parameters as given below 
\ba\label{p}
 &&g_m^{(A,B)}(z)=P_m^{(\alpha_m,\beta_m)}(z),\nonumber \\
&&\alpha_m = -A+1-m+\frac{B}{A-1+m},\qquad \beta_m= -A+1-m-\frac{B}{A-1+m},\nonumber \\
 &&m=1,2,3,....,.\qquad A>2,\quad (A-1)^2<B<(A-1)(A-1+m)\,,
\ea 
one obtains the rationally extended Eckart potentials,  
$V^{(-)}(r)$ (= $V_{A,B,ext}(r)$) isospectral to the potentials 
$V^{(+)}(r)$ with a bound state spectrum
\ba\label{bs}
&&E_\nu^{(+)}= E_\nu^{(-)} =  -(A-1+\nu )^2-\frac{B^2}{(A-1+\nu)^2}\,,
\qquad \nu=0,1,2,...,\nu_{max}\,,\nonumber \\
&&\sqrt{B}-A\le\nu_{max}<\sqrt{B}-A-1\,.
\ea    
The corresponding bound state eigenfunctions of $V^{(+)}(r)$ are 
\ba\label{wf1}
&&\psi_\nu^{(+)}(r) \propto  (z - 1)^{\frac{\alpha_\nu}{2}}
(z + 1)^{\frac{\beta_\nu}{2}}P_\nu^{(\alpha_\nu,\beta_\nu)}(z)\,,
\qquad \nu=0,1,2,....,\nu_{max}\,,\nonumber\\
&&\alpha_\nu = -A+1-\nu +\frac{B}{(A-1+\nu)}\,,\qquad 
\beta_\nu = -A+1-\nu -\frac{B}{(A-1+\nu)}\,,
\ea
and those of $V^{(-)}(r)$ are obtained by applying the operator 
$\hat{A}$ (as given by Eq. (\ref{op})) (in terms of $z$ variable)  
\ba\label{A}
\hat{A}&=&(1-z^2)\frac{d}{dz}+\frac{B}{A-1+m}-(A-1+m)z-(1-z^2)
\frac{\dot{g}_m^{(A,B)}(z)}{g_m^{(A,B)}(z)}\,, \nonumber\\
&=&(1-z^2)\frac{d}{dz}+\frac{B}{A-1}-(A-1)z
-\frac{2(m+\alpha_m )(m+\beta_m )}
{2m+\alpha_m+\beta_m}\frac{g_{m-1}^{(A+1,B)}(z)}{g_m^{(A,B)}(z)} \,,
\ea
on the bound state eigenfunctions of $V^{(+)}(r)$. The bound state 
eigenfunctions of $V^{(-)}(r)$ are then given by
\be\label{wf2}
\psi_\nu^{(-)}(r) \propto \frac{(z - 1)^{\frac{\alpha_\nu}{2}}
(z + 1)^{\frac{\beta_\nu}{2}}}{g_m^{(A,B)}(z)}y_n^{(A,B)}(z)\,,
\quad n=m+\nu-1,\quad \nu=0,1,2,....,\nu_{max}\,,
\ee
where $y_n^{(A,B)}(z)$ is some $n$th-degree polynomial in $z$, defined by
\ba\label{yn}
y_n^{(A,B)}(z)&=&\frac{2(\nu+\alpha_\nu )(\nu+\beta_\nu )}{
2\nu+\alpha_\nu+\beta_\nu}g_m^{(A,B)}(z)P_{\nu-1}^{(\alpha_\nu,\beta_\nu)}
(z)\nonumber \\
&-& \frac{2(m+\alpha_m )(m+\beta_m )}{2m+\alpha_m+\beta_m}
g_{m-1}^{(A+1,B)}(z)P_{\nu}^{(\alpha_\nu,\beta_\nu)}(z)\,,
\ea
and satisfying a second order differential equation 
\ba\label{deqn}
&\bigg[(1-z^2)&\frac{d^2}{dz^2}-\bigg \{\alpha_\nu -\beta_\nu +( \alpha_\nu + \beta_\nu +2)z + 2(1-z^2)\frac{\dot{g}_m^{(A,B)}(z)}{g_m^{(A,B)}(z)}\bigg \}\frac{d}{dz}\nonumber \\
&+&(\nu -1)(\alpha_\nu +\beta_\nu +\nu)-m(\alpha_m +\beta_m +m-1)\nonumber \\
&+&[\alpha_\nu -\beta_\nu+\alpha_m -\beta_m + (\alpha_\nu +\beta_\nu+\alpha_m +\beta_m )z]\frac{\dot{g}_m^{(A,B)}(z)}{g_m^{(A,B)}(z)}\bigg ]y_{m+\nu-1}^{(A,B)}(z)=0,\nonumber \\
&& \nu=0,1,2,....,\nu_{max}\,. 
\ea

 From the Eqs. (\ref{bs}), (\ref{wf1}) and (\ref{wf2}), we see that the 
energy eigenvalues of the extended shape invariant potentials 
$V^{(+)}(r)$ ($=V_{A+1,B}(r)$) and $V^{(-)}(r)$ ($=V_{A,B,ext}(r)$) are 
isospectral (broken SUSY), while the eigenfunctions are different.

\section{Scattering amplitude for extended shape invariant Eckart
potentials}

For obtaining the scattering amplitude of the new rationally-extended SI
Eckart potentials, we have to first obtain the scattering wave functions
for these potentials. 
On using  Eq. (\ref{yn}) and Eq. (\ref{wf2}) 
the bound state solutions for the rationally extended Eckart potentials 
are given by
\ba\label{wf3}
\psi_\nu^{(-)}(r)= \mbox{(Const.)}&\times & (z - 1)^{\frac{\alpha_\nu}{2}}(z + 1)^{\frac{\beta_\nu}{2}}
 \bigg[\frac{2(\nu+\alpha_\nu )(\nu+\beta_\nu )}{2\nu+\alpha_\nu+\beta_\nu}P_{\nu-1}^{(\alpha_\nu,\beta_\nu)}(z)-\nonumber \\
&&\frac{2(m+\alpha_m )(m+\beta_m )}{2m+\alpha_m+\beta_m}
\frac{g_{m-1}^{(A+1,B)}}{g_{m}^{(A,B)}}(z)
P_{\nu}^{(\alpha_\nu,\beta_\nu)}(z)\bigg]\,,
\ea
where $m=1,2,...,\quad \nu=0,1,2,....,\nu_{max}$\,.

 To get the scattering wave function for this rationally extended new 
Eckart potential, two modifications of the bound state wavefunctions 
have to be made \cite{ks}:
(i) The second solution of the Schrodinger equation must be retained - it has been discarded for bound state problems since it diverged asymptotically.
 (ii) Instead of the parameter $\nu $ labeling the number of nodes, one must use the wavenumber $k$ so that we get the asymptotic 
behavior in terms of $e^{\pm ikr}$ as $r\rightarrow \infty $.\\

The Jacobi polynomial in terms of the hypergeometric function $_2F_1$ 
is given by \cite{toi}
\be\label{hf}
P^{(\alpha_\nu , \beta_\nu )} _{\nu} (z) = (-1)^\nu 
\frac{\Gamma(\nu +\beta_\nu +1)}{ \nu! \Gamma (1+\beta_\nu )} 
{_2F_1}(\nu +\alpha_\nu +\beta_\nu +1, -\nu , 1+\beta_\nu ; \frac{1-z}{2})     
\ee
After considering the second solution of the Schr\"odinger  equation 
related to the bound state 
wave function ${\psi^{(-)}(r)}$ ( i.e, the second solution of the 
hypergeometric differential equation), the above equation becomes.
\ba\label{hf2}   
P^{(\alpha_\nu , \beta_\nu )} _{\nu} (\coth r)&=& (-1)^\nu\frac{\Gamma (\nu+\beta_\nu+1)}{ \nu ! \Gamma (1+\beta_\nu)} \left[C_1 \ \ {_2F_1}(\nu+\alpha_\nu +\beta_\nu+1 
,-\nu , 1+\beta_\nu ; \frac{1+\coth r}{2})\right. \nonumber \\
&+&\left. C_2 \bigg (\frac{1+\coth r}{2}\bigg )^{-\beta_\nu } 
{_2F_1}(\nu +\alpha_\nu+1 , -\nu -\beta_\nu  , 1- \beta_\nu ; \frac{1+\coth r}{2})\right], 
\ea
where $C_1$ and $C_2$ are constants assumed to be independent of 
$\nu$, $\alpha_\nu$, and $\beta_\nu$.
 
Considering the boundary condition, i.e as $r\rightarrow 0$ , $\psi_{\nu}^{(-)}(r) \rightarrow 0$,
and hence $C_2\rightarrow 0$, thus the allowed solution is
\be\label{alson}
P^{(\alpha_\nu , \beta_\nu )} _{\nu} (\coth r)=C_1(-1)^\nu\frac{\Gamma (\nu+\beta_\nu+1)}{ \nu ! \Gamma (1+\beta_\nu)}{_2F_1}(\nu+\alpha_\nu +\beta_\nu+1 
,-\nu , 1+\beta_\nu ; \frac{1+\coth r}{2}). 
\ee
Similarly by replacing $\nu$ by $\nu-1$ keeping in mind that $\alpha_\nu$ and $\beta_\nu$ are constants,
we get the expression for $P_{\nu-1}^{(\alpha_\nu,\beta_\nu)}\mbox{(coth r)}$. 

As $r\rightarrow \infty$ the new potential $V_{A,B,ext}(r)\rightarrow -2B$, hence let us define
\ba\label{sctt}
E_\nu^{(+)} (\mbox{or} E_\nu^{(-)})&-&V_{A,B,ext}(x\rightarrow\infty) = E_\nu^{(+)} (\mbox{or} E_\nu^{(-)})+2B\nonumber \\
 &=& -\left(A-1+\nu-\frac{B}{A-1+\nu}\right)^2 = k^2\qquad\mbox{(say)},
 \ea 
therefore  $\alpha_\nu = -ik$, where $k$ is wavenumber.\\
Also the polynomial $g_m^{(A,B)}(z)$ in terms of usual Jacobi polynomial is defined as
\ba\label{gm}
g_m^{(A,B)}(z)=P_m^{(\alpha_m,\beta_m)}(z)&=&\frac{\Gamma(\alpha_m +m+1)}{m!\Gamma(\alpha_m +\beta_m +m+1)}\nonumber\\
&\times &\sum_{p=0}^{m}\left( \begin{array}{clcr}
m \\
p
\end{array} \right)\frac{\Gamma(\alpha_m +\beta_m +m+p+1)}{\Gamma(\alpha_m +p+1)}\left(\frac{z-1}{2}\right)^p, \nonumber\\
\mbox{where}\quad m=1,2,3...,.
\ea
Similarly by replacing $m\rightarrow m-1$ and then changing $A\rightarrow A+1$, we get $g_{m-1}^{(A+1,B)}(z)$.\\

Note that the Jacobi polynomial as given by Eq. (\ref{alson}) is valid 
for any complex number $\nu \in\mathbb  C $, hence it is easy to see that 
the Schr\"odinger 
equation corresponding to Eq. (\ref{wf2}) holds for any complex number 
$\nu \in \mathbb{C} $ and the energy $ E_\nu^{+} (\mbox{or} E_\nu^{-})$ is real 
for $\alpha_\nu = -ik (k\in\mathbb{R})$. So, on replacing $\alpha_\nu$ by $-ik$, and using Eq.(\ref{gm}), a property 
related to the hypergeometric function \cite{toi} i.e.,\\
\ba\label{hf3}
{_2F_1}(a,b;c;z)&=&\frac{\Gamma(c)\Gamma(c-a-b)}{\Gamma(c-a)\Gamma(c-b)}{_2F_1}(a,b;a+b-c+1;1-z)
+(1-z)^{(c-a-b)}\nonumber \\
&&\times \frac{\Gamma(c)\Gamma(a+b-c)}{\Gamma(a)\Gamma(b)}
{_2F_1}(c-a,c-b;c-a-b+1;1-z)\,,
\ea
in Eq. (\ref{wf3}), then taking the asymptotic behavior of the 
wavefunction as $r\rightarrow \infty$,
and by using the fact that the hypergeometric function 
${_2F_1}(\alpha,\beta,\gamma; 0)\rightarrow 1$,
the scattering state wavefunction (\ref{wf3}) is given by
\be\label{scatt2}
\lim_{r\to\infty}\psi_{k}(r) = \mbox{(Const.)}
\times (S_{l=0}^m(k) e^{ikr} + e^{-ikr})\,.          
\ee
In this way, we find that the expression for the scattering amplitude 
$S_{l=0}^{m}(k)$ is given by
\ba\label{sctm2}
S_{l=0}^{m}(k) = S_{l=0}^{(+)}(k)\times \left[\frac{( A-ik+(m-1)
-\frac{B}{A+m-1})} {( A+ik+(m-1)-\frac{B}{A+m-1})} \right ]\,. 
\ea
Here $S_{l=0}^{(+)}(k)$ is the scattering amplitude for the potential 
$V^{(+)}(r)$ isospectral to $V^{(-)}(r)(=V_{A,B,ext}(r))$, given by 
\ba\label{sctt+}
S_{l=0}^{(+)}(k) = \frac{\Gamma(ik) \Gamma (-A+2-\frac{ik}{2}
-(B-\frac{k^2}{4})^\frac{1}{2})
\Gamma(A-1-\frac{ik}{2}- (B-\frac{k^2}{4})^\frac{1}{2})}
{\Gamma(-ik)\Gamma (-A+2+\frac{ik}{2}-(B-\frac{k^2}{4})^\frac{1}{2})
\Gamma (A-1+\frac{ik}{2}- (B-\frac{k^2}{4})^\frac{1}{2})}\,.
\ea
After simplifying $S_{l=0}^{(+)}(k)$, Eq. (\ref{sctm2}) is written as 
\ba\label{sctt1}
S_{l=0}^{m}(k) = S_{l=0}^{Eckart}(k)\times 
\left[\frac{(A+ik-1-\frac{B}{A-1})( A-ik+(m-1)-\frac{B}{A+m-1})} 
{( A-ik-1-\frac{B}{A-1}) ( A+ik+(m-1)-\frac{B}{A+m-1})} \right ]\,, 
\ea
where $S_{l=0}^{Eckart}(k)$ is the scattering amplitude for the 
conventional Eckart potential $(V^{(A,B)}(r))$, given by \cite{ks}
\ba 
S_{l=0}^{Eckart}(k) = \frac{\Gamma (ik) \Gamma (A-\frac{ik}{2}+(B-\frac{k^2}{4})^\frac{1}{2})\Gamma (A-\frac{ik}{2}- (B-\frac{k^2}{4})^\frac{1}{2})}{\Gamma (-ik)\Gamma (A+\frac{ik}{2}+(B-\frac{k^2}{4})^\frac{1}{2})\Gamma (A+\frac{ik}{2}- (B-\frac{k^2}{4})^\frac{1}{2})} a_k \nonumber
\ea
here $a_k$ is a phase factor.

As a check on our calculations, by starting from Eq. (\ref{wf3}),  
we have also explicitly calculated the scattering amplitudes in the 
special cases of $m=1, 2$ and $3$ and checked that we get the same 
expressions as obtained from Eq. (\ref{sctt1}) with $m=1,2$ and $3$ 
respectively. Further, in the 
limit $m=0$, the scattering amplitude as given by Eq. (\ref{sctt1}) reduces
to $S_{l=0}^{Eckart}(k)$, providing a further check on our  calculations.

\section{Summary and discussion}  

Recently the bound state eigenfunction of the rationally-extended Eckart 
potential, which is isospectral to the conventional Eckart potential, has
been written by Quesne in terms of some new polynomials. These new 
polynomials are not EOPs, but are related to the usual Jacobi polynomials. 
In this paper we have considered the scattering problem for the one parameter 
family of such rationally-extended Eckart potentials and have obtained 
the scattering amplitude for these new rationally extended Eckart 
potentials. We have shown that the scattering amplitude for the general 
$m$th case ($m$ is the order of the new polynomials) is related to the 
scattering amplitude for the conventional Eckart potential. In the special
case of $m=0$, as expected, we recover the scattering amplitude for the usual 
Eckart potential.  

{\bf Acknowledgment}

One of us (RKY) acknowledges financial support from UGC under the FIP Scheme.

\end{document}